\documentclass[journal,letterpaper]{IEEEtran}
\IEEEoverridecommandlockouts

\usepackage{fancyhdr}
\usepackage{cite}
\usepackage{graphicx}
\usepackage{psfrag}
\usepackage{subfigure}
\usepackage{url}
\usepackage{balance}
\usepackage{stfloats}
\usepackage{amsmath,amsthm,amssymb}

\DeclareMathOperator*{\argmin}{arg\,min}
\usepackage{array}
\usepackage{fancyhdr}
\usepackage{float}
\usepackage{epsfig}
\usepackage{color}
\usepackage[nomain,acronym,toc]{glossaries}
\usepackage{algorithm,algorithmic}
\usepackage{epstopdf}
\usepackage{multirow,multicol}

\newacronym{urllc}{URLLC}{ultra-reliable low-latency communication}
\newacronym{itu}{ITU}{international telecommunication union}
\newacronym{ccdf}{CCDF}{complementary cumulative distribution function}
\newacronym{cdf}{CDF}{cumulative distribution function}
\newacronym{snr}{SNR}{signal-to-noise ratio}
\newacronym{ofdma}{OFDMA}{orthogonal frequency-division multiple-access}
\newacronym{se}{SE}{spectral efficiency}
\newacronym{fdma}{FDMA}{frequency-division multiple-access}
\newacronym{csit}{CSIT}{channel-state-information at the transmitter}
\newacronym{mac}{MAC}{medium-access control}
\newacronym{tti}{TTI}{transmission time interval}
\newacronym{harq}{HARQ}{hybrid automatic repeat request}
\newacronym{isnr}{iSNR}{instantaneous signal-to-noise-ratio}

\definecolor{sblue}{RGB}{0,51,120}
\definecolor{sred}{RGB}{200,51,130}

\def\BibTeX{{\rm B\kern-.05em{\sc i\kern-.025em b}\kern-.08em
		T\kern-.1667em\lower.7ex\hbox{E}\kern-.125emX}}
\begin{document}

	\title{Block-Weighted Lasso for Joint Optimization of Memory Depth and Kernels in Wideband DPD
	}
	\author{Jinfei Wang, Yi Ma, Fei Tong, and Ziming He
			\thanks{\hrule\vspace{5pt} Jinfei Wang and Yi Ma are with the Institute for Communication Systems~(ICS), University of Surrey, Guildford, UK, GU2 7XH. Emails: (jinfei.wang, y.ma)@surrey.ac.uk.}			
			\thanks{Fei Tong and Ziming He are with the Samsung Cambridge Solution Centre, Cambridge, UK, CB4 0AE. Emails: (f.tong, ziming.he)@samsung.com}
			\thanks{This work was funded by Samsung-Academy Global Collaborative Research Programme. Our sincere appreciation to Zern C. Tay and Jacob Sharpe for their outstanding support on testbed measurements.} 
}


	\maketitle

\begin{abstract}

The optimizations of both memory depth and kernel functions are critical for wideband digital pre-distortion (DPD).
However, the memory depth is usually determined via exhaustive search over a wide range for the sake of linearization optimality, followed by the kernel selection of each memory depth, yielding excessive computational cost.
In this letter, we aim to provide an efficient solution that jointly optimizes the memory depth and kernels while preserving reasonable linearization performance.
Specifically, we propose to formulate this optimization as a block-weighted least absolute shrinkage and selection operator (Lasso) problem, where kernels are assigned regularization weights based on their polynomial orders.
Then, a block coordinate descent algorithm is introduced to solve the block-weighted Lasso problem.
Measurement results on a generalized memory polynomial (GMP) model demonstrates that our proposed solution reduces memory depth by {$31.6\%$} and kernel count by $85\%$ compared to the full GMP, while achieving $–46.4$~dB error vector magnitude (EVM) for signals of $80$~MHz bandwidth. 
In addition, the proposed solution outperforms both the full GMP and the GMP pruned by standard Lasso by at least $0.7$~dB in EVM.

\end{abstract}


\section{Introduction}
\IEEEPARstart{A}{s wireless communication} bandwidth continues to expand, the strong memory effects exhibited by power amplifiers (PAs) necessitates the optimizations of memory depth and kernel functions in digital pre-distortion (DPD) \cite{Masterson2022,9955534,10245788,Akram2023}. 
{In the literature, memory depth is often first determined via exhaustive search over a wide range of depths to ensure the linearization performance (e.g., \cite{BenMabrouk2015,Morgan2006},}) followed by kernel selection to prune the full DPD model~\cite{9730199,MarquesValderrama2024,4547349}.
While effective, this two-step process incurs excessive computational cost.

In this letter, we aim to provide an efficient solution that jointly optimizes the memory depth and kernels while preserving reasonable linearization performance.
It is mathematically challenging to explicitly incorporate the memory depth in the optimization objective.
To address this issue, we transform the problem into a kernel selection problem but with the additional goal of encouraging selected kernels to collectively yield a reduced memory depth.
In the literature, kernel selection is usually formulated as a standard least absolute shrinkage and selection operator (Lasso) problem~\cite{9730199,MarquesValderrama2024,4547349}.
However, our measurement results (see Section~\ref{secIII}) reveal that standard Lasso tends to retain the maximum available memory depth, failing to meet our design objective. 
This behavior may arise because standard Lasso applies uniform regularization weight across all kernels, despite their behavior could vary depending on their polynomial orders.

To overcome this limitation, we propose to formulate this joint optimization as a block-weighted Lasso problem.
The principle is to assign different regularization weights to kernels of different polynomial order, naturally grouping kernels of identical polynomial orders into blocks.
To solve this problem, we leverage the widely used block coordinate descent (BCD) algorithm, which iteratively updates the kernel coefficients of each block \cite{Xu2025,Tseng2001}.

Measurement results on a generalized memory polynomial (GMP) model demonstrates that our proposed solution reduces memory depth by {$31.6\%$ (from $19$ to $13$)} and kernel count by $85\%$ compared to the full GMP (from $300$ to $46$), while achieving $–46.4$~dB error vector magnitude (EVM) for orthogonal frequency division multiplexing (OFDM) signals of $80$~MHz. 
In comparison, the full GMP and GMP pruned by standard Lasso achieve $-45.1$~dB and $-45.7$~dB of EVM, respectively.
This confirms the effectiveness of our proposed solution.

%
%

\section{Block-Weighted Lasso for\\ Joint Memory Depth and Kernel Optimization}

\subsection{Problem Statement and State-of-The-Art}

Our objective is to jointly determine the memory depth and selected kernels for the DPD model.
It is mathematically challenging to explicitly incorporate the memory depth in the optimization objective.
To address this issue, we transform the objective into a kernel selection problem, where the selected kernels are expected to collectively yield an optimized memory depth smaller than the maximum memory depth available.

Kernel selection has been extensively studied to simplify DPD models.
In state-of-the-art, it is usually formulated as a standard Lasso problem \cite{10245788, Morgan2006, Abdelhafiz2018}.
However, our measurement results indicate that the standard Lasso tends to retain full available memory depth, thereby contradicting our goal.
This behavior may arise because standard Lasso applies uniform regularization across all kernels, despite their behavior could vary depending on their polynomial orders.

To facilitate our study, we present the mathematical formulation of standard Lasso problem for the GMP model \cite{10245788, Morgan2006, Abdelhafiz2018}.
But note that our discussion can be extended to other DPD models that support Lasso-based kernel selection.
Consider an OFDM signal $\mathbf{s}\in\mathbb{C}^{N\times1}$.
Assume its label signal for DPD is known (denoted by $\mathbf{x}\in\mathbb{C}^{N\times1}$ ), e.g., through iterative learning control (ILC) \cite{7522645,Xia2021,10018540,Wang2025}.
The GMP describes the function from $\mathbf{s}$ to $\mathbf{x}$ as follows \cite{He2022}:
\begin{IEEEeqnarray}{rl}
x(n) =& \sum_{k\in\mathcal{K}_a}\sum_{l\in\mathcal{L}_a}a_{kl}s(n-l)|s(n-l)|^k\IEEEnonumber\\
	&+\sum_{k\in\mathcal{K}_b}\sum_{l\in\mathcal{L}_b}\sum_{m\in\mathcal{M}_b}b_{klm}s(n-l)|s(n-l-m)|^k\IEEEnonumber\\
	&+\sum_{k\in\mathcal{K}_c}\sum_{l\in\mathcal{L}_c}\sum_{m\in\mathcal{M}_c}c_{klm}s(n-l)|s(n-l+m)|^k,\label{eq01}
\end{IEEEeqnarray}
where $s(n)$ stands for the $n$-th element of $\mathbf{s}$, $x(n)$ for the $n$-th element of $\mathbf{x}$, $\mathcal{K}_a$, $\mathcal{L}_a$ and $a_{kl}$ for the order/memory index arrays and coefficients of aligned terms, $\mathcal{K}_b$, $\mathcal{L}_b$, $\mathcal{M}_b$ and $b_{klm}$ for the order/memory/cross-memory index arrays and coefficients of lagging cross terms, $\mathcal{K}_c$, $\mathcal{L}_c$, $\mathcal{M}_c$ and $c_{klm}$ for the order/memory/cross-memory index arrays and coefficients of leading cross terms.  

{
Each term in \eqref{eq01} is a kernel function of the GMP model.
Before kernel selection, the full GMP model is considered, where the index arrays span the entire range up to a given memory depth $L$, polynomial order $K$, the lagging depth $M_b$:
\begin{IEEEeqnarray}{rl}
\mathcal{L}_a,\mathcal{L}_b,\mathcal{L}_c&=\{0,1,\cdots,L\},\\
\mathcal{K}_a,\mathcal{K}_b,\mathcal{K}_c&=\{0,2,\cdots,K-1\},\\
\mathcal{M}_b&=\{1,\cdots,M_b\}.
\end{IEEEeqnarray}
The inclusion of leading cross-terms is determined based on design decisions, as discussed in \cite{Morgan2006}.
}

Denote the total kernel count by $P$.
To facilitate our study, we represent the coefficients of all kernels as a vector $\boldsymbol{\omega}$, where $\boldsymbol{\omega}\in\mathbb{C}^{P\times1}$.
Then, the standard Lasso problem is given by:
\begin{equation}\label{eqLasso}
\boldsymbol{\omega}^\star = \argmin_{\boldsymbol{\omega}} \|\mathbf{x}-\mathbf{S}\boldsymbol{\omega}\|^2_2+\lambda\|\boldsymbol{\omega}\|_1,   
\end{equation}
where $\|\cdot\|_1$ stands for the $\l_1$ norm, $\|\cdot\|_2$ for Euclidean norm, $\mathbf{S}\in\mathbb{C}^{N\times P}$ for the kernel matrix (see \cite[(25)]{Morgan2006}), $\lambda\in\mathbb{R}^+$ for the regularization weight. 
A larger $\lambda$ typically yields a sparser $\boldsymbol{\omega}^\star$ \cite{Tibshirani1996}.

\subsection{Block-Weighted Lasso Problem}

The standard Lasso has proven successful in reducing the kernel count \cite{10245788,9825005}. 
But specifically for our objective, measurement results show that the standard Lasso tends to retain the maximum available memory depth (see Section.~\ref{secIII}).
Our hypothetic explanation is that this phenomenon arises from the uniform regularization weight applied across all kernels, since kernels of different polynomial orders can exhibit distinct behaviors for their different amplitude magnitudes and portion of contribution to the output signal.
Consequently, directly applying the stand Lasso cannot meet our objective.

In light of this hypothesis, we propose to assign different regularization weights to the $\l_1$-norm of each kernel based on the polynomial order.
This naturally breaks down the kernels into blocks according to their polynomial orders,  and yields the following block-weighted Lasso formulation:
\begin{equation}\label{eqWLasso}
\boldsymbol{\omega}^\star = \argmin_{\boldsymbol{\omega}} \|\mathbf{x}-\mathbf{S}\boldsymbol{\omega}\|^2_2+\sum_{k\in\mathcal{K}}\lambda_k\|\boldsymbol{\omega}_k\|_1,   
\end{equation} 
where $\lambda_k$ and $\boldsymbol{\omega}_k$ stand for the regularization weight and coefficients for the kernels with polynomial order of $(k+1)$.
In Section~\ref{secIV}, the setting of $\lambda_k$ will be discussed in detail.


\subsection{BCD Algorithm for Block-Weighted Lasso}

{
\begin{algorithm}[t]
\caption{BCD Algorithm for Block-Weighted Lasso}
\begin{algorithmic}[1]\label{agthm1}
\renewcommand{\algorithmicrequire}{\textbf{Input:}}
\renewcommand{\algorithmicensure}{\textbf{Output:}}
\REQUIRE OFDM signal $\mathbf{s}$, DPD signal $\mathbf{x}$, iteration number $R$, memory depth $L$, polynomial order $K$, regularization weights $\lambda_0,\cdots,\lambda_{K-1}$;
\ENSURE Coefficient vector $\boldsymbol{\omega}$;
\STATE \textbf{initialization}: $\boldsymbol{\omega}_0=\mathbf{0},\cdots,\boldsymbol{\omega}_{K-1}=\mathbf{0}$; 

\FOR{$r=1:R$}
\FOR{$k=0:2:K-1$}
\STATE Calculate $\tilde{\mathbf{x}}_k$ with \eqref{eqWLassoBlock2};
\STATE Update $\boldsymbol{\omega}_k$ by solving \eqref{eqWLassoBlock};
\ENDFOR
\ENDFOR
\STATE $\boldsymbol{\omega}\leftarrow[\boldsymbol{\omega}_{1};\boldsymbol{\omega}_{3};\cdots;\boldsymbol{\omega}_{K}]$;
\RETURN $\boldsymbol{\omega}$. 
\end{algorithmic} 

\end{algorithm}
	
}

The block structure of \eqref{eqWLasso} motivates us to leverage the widely used BCD algorithm as solution \cite{Xu2025,Tseng2001}.
The idea of BCD algorithm is to update each block of coefficients sequentially, one at a time, while keeping the others fixed \cite{Xu2025}.
When updating the block of order $(k+1)$, the update aims to solve the following problem:
\begin{equation}\label{eqWLassoBlock}
\boldsymbol{\omega}_k^\star = \argmin_{\boldsymbol{\omega}_k} \|\tilde{\mathbf{x}}_k-\mathbf{S}_k\boldsymbol{\omega}_k\|^2_2+\lambda_k\|\boldsymbol{\omega}_k\|_1,   
\end{equation} 
where
\begin{equation}\label{eqWLassoBlock2}
\tilde{\mathbf{x}}_k \triangleq \mathbf{x}-\sum_{g\neq k}\mathbf{S}_g\boldsymbol{\omega}_g,   
\end{equation} 
and $\mathbf{S}_k$ stands for the sub-matrix of $\mathbf{S}$ corresponding to the polynomial order of $(k+1)$.
This reduces the problem to a standard Lasso for each block.
Moreover, recognizing that the sequential updates can introduce sub-optimality in regression, an iterative refinement procedure is introduced to enhance the regression performance \cite{Xu2025}.
The adopted BCD algorithm is summarized in \textbf{Algorithm~\ref{agthm1}}.

A critical aspect of the BCD algorithm is the convergence behavior \cite{Xu2025,Tseng2001}.
Due to space limit, the convergence behavior of \textbf{Algorithm~\ref{agthm1}} is demonstrated empirically through measurement results later in Section.~\ref{secIII}, and the convergence analysis is left in future studies.

\section{Measurement Results and Discussion}\label{secIII}

In this section, the performance of the proposed block-weighted Lasso is demonstrated through measurement results for Wi-Fi OFDM signals of $80$~MHz bandwidth, including two experiments: \textit{Experiment~$1$} shows convergence behavior in normalized mean square error (NMSE) and the kernel selection performance in memory depth and kernel count; \textit{Experiment~$2$} shows the linearization performance in EVM.

The testbed consists of a Rhode \& Schwarz SMM100A vector signal generator, a Rhode \& Schwarz FSW vector signal analyzer, and a Qorvo QM45500 PA in low-power mode. 
The carrier frequency is $5.32$~GHz. 
Both digital-analogue converter (DAC) and analogue-digital converter (ADC) resolution are $12$~bits; the DAC rate is $480$~MHz; the ADC rate is $640$~MHz.
The DPD training signal $\mathbf{x}$ is obtained via ILC \cite{7522645}.
The adopted solution to standard Lasso is the iterated ridge regression for both step~$5$ in \textbf{Algorithm~\ref{agthm1}} and the standard Lasso baseline in appreciation of its {computational efficiency and stability} \cite{Yang2022,CS542B_UCB}.
{The GMP structure considered has {$L=19$}, $K=15$, $M_b=1$, $\mathcal{M}_c=\emptyset$, i.e., the total kernel count is $300$.}

\begin{table}[t!]
	\center
	\caption{Convergence Behavior of \textbf{Algorithm~\ref{agthm1}}}
	\label{tabIII}
	\begin{tabular}{c||c|c|c|c|c}
		\hline
		Iteration & $1$ & $2$ & $3$ & $4$ & $5$ \\
		\hline
		NMSE (dB) & $-34.28$  & $-37.70$ & $-37.75$ & $-37.76$ & $-37.91$ \\
		\hline
		Iteration & $6$ & $7$ & $8$ & $9$ & $10$\\
		\hline
		NMSE (dB) & $-36.87$  & $-37.75$ & $-37.86$ & $-37.86$ & $-36.99$ \\
		\hline
	\end{tabular}
\end{table}

\begin{figure}[t]
	\centering
	\includegraphics[scale=0.5]{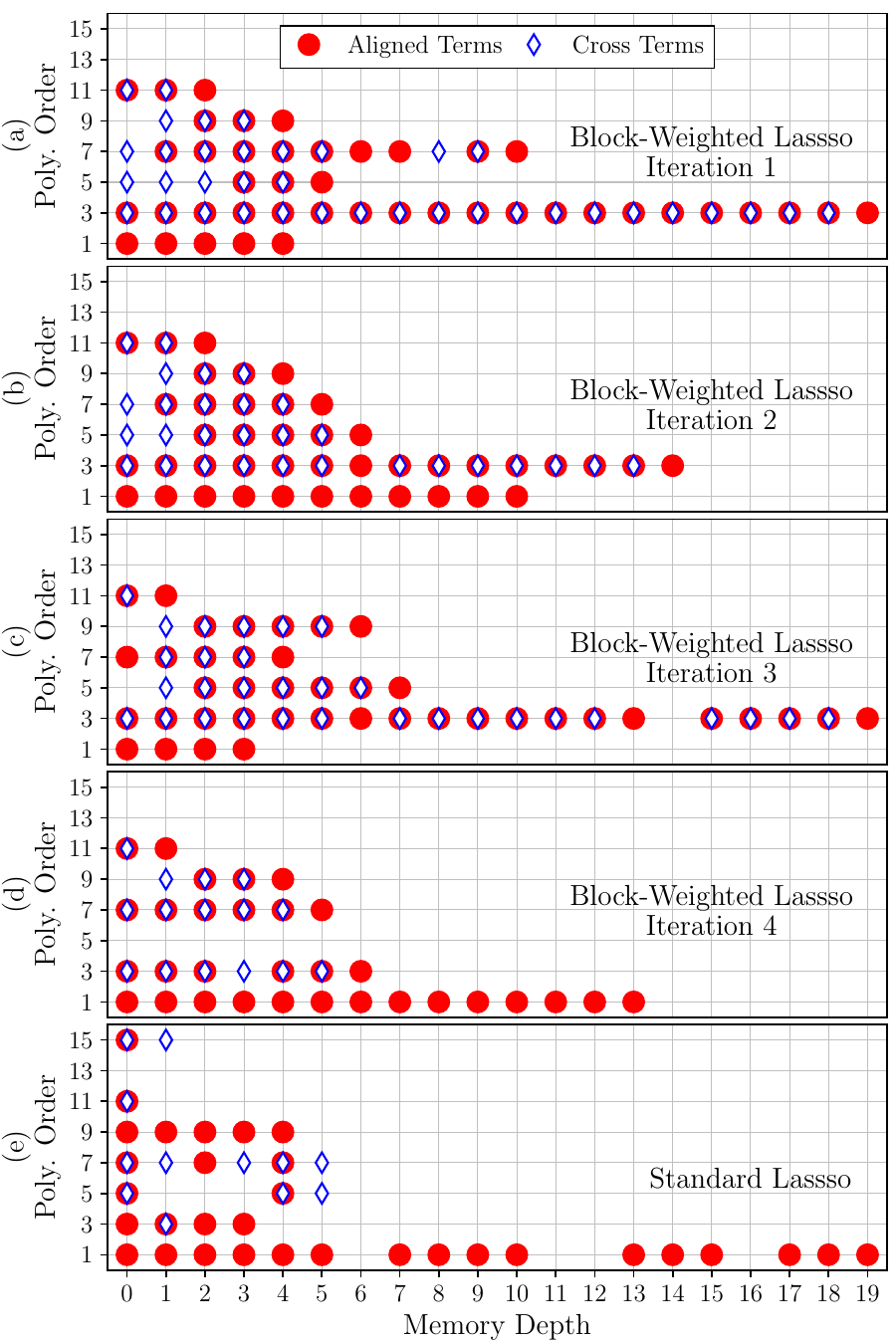}
	\caption{Memory depth and kernels selected by the block-weighted Lasso compared to the standard Lasso baseline.}
	\label{fig2}
\end{figure}

{We assign higher regularization weight to kernels of higher polynomial orders (i.e., higher sparsity) based on the understandings that lower-order kernels (e.g., $7\sim9$) make most contributions to the DPD output \cite{Morgan2006,9583897}.}
Specifically, we progressively increase $\lambda$: $\lambda_0=10^{-4}$ for linear kernels; $\lambda_k=1.35\lambda_{k-2}$ when $2\leq k\leq8$; $\lambda_k=2\lambda_{k-2}$ when $10\leq k\leq14$.
The same rule is applied to the zero threshold parameter for iterated ridge regression, which starts from $0.17$ for linear kernels.

\textit{Experiment~1:} This experiment aims to demonstrate the convergence behavior of the proposed solution in NMSE as well as its selection of memory depth and kernels.
Table.~\ref{tabIII} shows the regression NMSE with respect to the iteration.
From the second iteration, the NMSE sticks around $-37.7$~dB with a fluctuation less than $0.9$~dB. 
This confirms the convergence of the proposed solution in terms of regression accuracy.
For reference, the regression NMSE of full GMP using least-square (LS) is $-37.60$~dB, indicating that the proposed solution achieves reasonable regression accuracy.

In addition, Fig.~\ref{fig2}(a)$\sim$(d) demonstrates the selected kernels of the proposed solution from iteration $1$ to $4$.
It is demonstrated that at the $4$-th iteration, the proposed solution successfully reduced the memory depth {from $19$ to $13$ ($31.6\%$ reduction)} and the kernel count from $300$ to $46$ ($85\%$ reduction).
In comparison, Fig.~\ref{fig2}(e) shows the kernels selected by standard Lasso when the kernel count is $44$.
Despite the similar kernel count, standard Lasso still picks the maximum memory depth, as emphasized before.

Overall, the results demonstrate that the proposed solution effectively achieves joint optimization of memory depth and kernels.


\subsection*{Experiment~2:}
This experiment aims to demonstrate the linearization performance of the proposed solution in terms of EVM in comparison to LS solution of full GMP and GMP pruned by standard Lasso.
To mitigate potential coefficient bias introduced by $\l_1$ regularization, LS fitting is applied to refine the coefficients of the selected kernels for the proposed solution and standard Lasso baseline. 
In Table~\ref{tabII}, results obtained after LS refinement are labeled by ‘R’ (refined), while those without refinement are labeled ‘NR’ (non-refined).

Table.~\ref{tabII} demonstrates that the proposed solution with LS refinement achieves the best EVM of $-46.4$~dB, outperforming the standard Lasso for $0.7$~dB and the LS of full GMP for $1.3$~dB.
These results confirm that the proposed solution provides reasonable linearization performance. 

When combined with the results from \textit{Experiment~1}, it can be concluded that the proposed block-weighted Lasso effectively fulfills its design objectives.

\begin{table}[t!]
	\center
	\caption{Linearization Performance of Block-Weighted Lasso}
	\label{tabII}
	\begin{tabular}{c||c|c|c|c|c|c}
		\hline
		\multirow{3}{*}{ } & w/o & \multirow{3}{*}{LS} & \multicolumn{2}{c|}{standard} & \multicolumn{2}{c}{block-weighted}\\
		& DPD &  & \multicolumn{2}{c|}{Lasso} & \multicolumn{2}{c}{Lasso}\\
		\cline{4-7}
		&  &  &  NR & R  & NR  & R \\
		\hline
		EVM (dB) & -37.6  & -45.1  & -45.7 & -44.7 & -45.5 & \textbf{-46.4}  \\
		\hline
	\end{tabular}
\end{table}

\section{Conclusion}\label{secIV}

In this letter, a block-weighted Lasso problem has been proposed to jointly optimize the memory depth and kernels for DPD models, where the regularization weights of kernels are assigned according to their polynomial orders; a BCD algorithm has been introduced to solve the proposed problem.
Measurement results on a GMP model have demonstrated that our proposed solution reduces memory depth by {$31.6\%$} and kernel count by $85\%$, while achieving EVM of $–46.4$~dB for OFDM signals of $80$~MHz.

\balance


	
	
	\newpage
	\bibliographystyle{IEEEtran}
	\bibliography{bib/digitalPreDistortion,bib/GroupPaper}	
	

	%
	
\end{document}